\documentclass[12pt,a4paper]{article}

\usepackage{graphicx}
\usepackage[top=2cm, bottom=2cm, left=2cm, right=2cm, includeheadfoot]{geometry}
\usepackage{microtype}
\usepackage{amsmath}
\usepackage{xcolor}
\usepackage[normalem]{ulem}   
\usepackage{url}

\usepackage{listings}
\lstdefinestyle{pycode}{
  language=Python,
  basicstyle=\small\ttfamily,
  keywordstyle=\color{blue!60!black}\bfseries,
  commentstyle=\color{black!55}\itshape,
  stringstyle=\color{green!40!black},
  emph={LinearOperator,cg},
  emphstyle=\color{purple!60!black},
  numbers=none,
  frame=single,
  framerule=0.4pt,
  rulecolor=\color{black!25},
  backgroundcolor=\color{black!4},
  xleftmargin=1.5em,
  xrightmargin=1.5em,
  showstringspaces=false,
  columns=fullflexible,
  keepspaces=true,
  breaklines=true,
  captionpos=t,
  belowcaptionskip=6pt,
}
\usepackage{booktabs}
\usepackage{authblk}
\usepackage{cleveref}
\usepackage{setspace}
\usepackage{caption}
 
\crefname{equation}{Eq.}{Eqs.}
\Crefname{equation}{Equation}{Equations}
\crefrangelabelformat{equation}{#3(#1)--(#2)#4}
\crefname{figure}{Fig.}{Figs.}
\Crefname{figure}{Figure}{Figures}
\crefname{table}{Table}{Tables}
\Crefname{table}{Table}{Tables}
\crefname{lstlisting}{Listing}{Listings}
\Crefname{lstlisting}{Listing}{Listings}
\usepackage[backend=biber,style=phys,biblabel=superscript,articletitle=true,
            pageranges=false,sorting=none,maxbibnames=5,minbibnames=5,
            doi=false,url=false,isbn=false]{biblatex}
\DeclareCiteCommand{\cite}
  {\usebibmacro{prenote}}
  {Ref.~\printtext[bibhyperref]{\printfield{labelnumber}}}
  {\multicitedelim}
  {\usebibmacro{postnote}}
\title{PyCDFT: A Python-scriptable library for analytical evaluation of orbital conceptual density (matrix) functional theory}
\author[1*]{Bin Wang}
\author[1]{Paul Geerlings}
\author[2]{Paul W. Ayers}
\author[1]{Frank De Proft}
\affil[1]{\textit{Research Group of General Chemistry (ALGC), Vrije Universiteit Brussel (VUB), Pleinlaan 2, B-1050 Brussels, Belgium. E-mail:} bin.wang@vub.be}
\affil[2]{\textit{Department of Chemistry and Chemical Biology, McMaster University, Hamilton, ON, Canada.}}
\date{}

\begin{document}

\maketitle

\begin{center}
\section*{Abstract}
\end{center}

Conceptual density functional theory (CDFT) defines chemical reactivity descriptors as derivatives of the electronic energy with respect to the number of electrons and the external potential or combinations thereof; in practice, however, these derivatives are almost always replaced by finite-difference and frontier-orbital approximations, and no common software exists for their analytical evaluation. We present \textsc{PyCDFT}, to our knowledge the first standardized, open-source, and scriptable code that analytically computes the descriptors of conceptual density (matrix) functional theory for ground and excited states up to second order, i.e., the orbital hardness, the Fukui function and Fukui matrix, and the linear response function, from a single converged mean-field wavefunction imported from essentially any electronic-structure package. The descriptors are obtained from matrix-free, preconditioned Krylov-subspace solutions of the coupled-perturbed self-consistent-field equations for both spin-unpolarized and spin-polarized references, with the work distributed over MPI ranks and all data flowing through a single HDF5 checkpoint file that supports restart, post-processing, and the export of real-space descriptors as cube files for visualization. The design delegates integrals, grids, and exchange--correlation kernels to PySCF and Libxc, keeping the library compact, user-friendly, and interoperable. Worked examples on the ground state of H$_2$O and a broken-symmetry $\Delta$SCF excited state of NH$_3$ show that a complete second-order CDFT analysis requires only a few lines of user code.

\section{Introduction}

Over more than four decades, conceptual density functional theory (CDFT),\autocite{Parr1989,Chermette1999,Geerlings2003,Ayers2005,Gazquez2008,Liu2009,Geerlings2020,Liu2022,Liu2024} also referred to as density functional reactivity theory (DFRT), has matured into a firmly rooted framework for the study of chemical reactivity. Its basic idea, going back to Parr and co-workers in the late 1970s and early 1980s,\autocite{Parr1978} is that concepts that chemists had long used intuitively can be given precise, computable definitions as derivatives (response functions) of the energy functional $E[N,v]$ with respect to the number of electrons $N$ and the external potential $v(\mathbf{r})$: the electronic chemical potential $\mu$\autocite{Mulliken1934,Iczkowski1961,Parr1978} and the density $\rho(\mathbf{r})$\autocite{Parr1989} appear at first order, followed at second order by the chemical hardness $\eta$,\autocite{Pearson1963,Parr1983,Parr1991,Pearson1993,Pearson1997} the Fukui function $f(\mathbf{r})$,\autocite{Parr1984,Yang1984} and the linear response function (LRF) $\chi(\mathbf{r},\mathbf{r}')$,\autocite{Parr1989,Sablon2010,Boisdenghien2013,Geerlings2014,Fias2022,Wang2023} while higher-order response functions\autocite{Geerlings2008} follow in the same way. A reaction is thereby viewed as the response of a molecule to the changes $\Delta N$ and $\Delta v(\mathbf{r})$ imposed by its reaction partner. In applications, these derivatives are seldom computed from their definitions. The $N$-derivatives are often estimated by finite differences involving the ionized systems ($\mu\approx-(I+A)/2$, $\eta\approx I-A$), the Fukui function is on the other hand obtained from $N\pm1$ density differences or identified with a frontier orbital density,\autocite{Parr1984} and atomic resolution is introduced through a partitioning scheme.\autocite{Yang1986Mortier,Wang2019} This toolbox has carried most applied CDFT work, but each ingredient has known weaknesses: the calculations on anions in the finite difference approximation can be difficult to converge and are prone to spin contamination and delocalization error,\autocite{Cohen2008,Cohen2012,Wang2024hardness,Wang2024} the frontier orbital picture ignores orbital relaxation and fails for (quasi-)degenerate frontier orbitals,\autocite{Bultinck2014} and the outcome may change qualitatively with the choice of approximation or partitioning method.\autocite{Wang2019} For the LRF, the difficulties become prohibitive: a two-point kernel cannot be obtained by finite differences, as this would require perturbing $v(\mathbf{r})$ numerically and independently at every point in space, which is prohibitively expensive, and for a long time $\chi(\mathbf{r},\mathbf{r}')$ was accessible only through simplified models or truncated expressions such as the independent particle approximation (IPA).\autocite{Sablon2010,Sablon2010deloc,Sablon2010indres} The decisive first step was the analytical evaluation of the LRF from coupled-perturbed self-consistent-field (CP-SCF) theory,\autocite{Yang2012,Geerlings2014} which yields the kernel exactly for a given density functional approximation;\autocite{Wang2023} on this basis the LRF has been applied to problems ranging from atomic shell structure\autocite{Boisdenghien2013} and substituent effects to electron delocalization\autocite{Sablon2010deloc,Sablon2010indres} and the nearsightedness of electronic matter,\autocite{Fias2017} with quantitatively reliable and readily interpretable results.

The analytical route also pays off beyond reactivity analysis itself. In the development of density functionals,\autocite{Mei2021} the deviation between the analytical and finite-difference Fukui functions makes the delocalization error of an approximate functional directly visible,\autocite{Wang2024} and the analytically evaluated hardness permits an explicit test of the hardness condition that the exact functional must obey.\autocite{Wang2024hardness} Moreover, the same derivative framework is not restricted to the perturbation ($N$ and $v(\mathbf{r})$) exerted by an approaching reagent. Treating the nuclear charges as continuous variables connects CDFT to alchemical perturbation theory with a rational exploration of chemical compound space coming within range.\autocite{Sheppard2010,Fias2017,Balawender2018}. Reactivity indices have recently been reformulated for molecules under external electric\autocite{Clarys2021} and magnetic fields,\autocite{Francotte2022,Teale2024} mechanical forces,\autocite{Bettens2019,Bettens2022} and isotropic pressure\autocite{Eeckhoudt2022,Eeckhoudt2024} or chemical constraints.\autocite{Kaduk2012} In all of these settings, finite-difference protocols become increasingly challenging or sometimes ill-defined, whereas an analytical formulation provides a single systematic starting point and leaves room for perturbations that have not been considered yet. Meanwhile, the theory continues to grow in directions where analytical evaluation is unavoidable: generalizing the external potential to the one-electron integrals $h_{pq}$ is the gate to conceptual density matrix functional theory (CDMFT) extending the theory from atoms to bonds;\autocite{Wang2025} varying individual occupation numbers defines orbital-resolved CDFT,\autocite{Liu1997,Wang2026} and descriptors under external fields\autocite{Franco2024} and higher-order response functions\autocite{Geerlings2008,Hoffmann2026} are under active investigation. All of these developments, however, run into the same practical bottlenecks: (1) there is no common, openly available code for analytical CDFT,\autocite{Guerrero2026} so the CP-SCF machinery is re-implemented separately for each study; (2) the notation for the response functions varies considerably across the literature, making the published expressions difficult to compare and combine; and (3) no single reference work explains how the analytical framework should be implemented and carried through to its most general orbital and matrix extensions. In the present work we therefore introduce \textsc{PyCDFT}, an open, Python-scriptable library that analytically computes the orbital CD(M)FT descriptors up to second order from a single converged wavefunction. Sec.~\ref{sec:fundamentals} collects the working equations, Sec.~\ref{sec:development} describes the structure and implementation of the library, Sec.~\ref{sec:examples} illustrates its use, and Sec.~\ref{sec:conclusions} closes with an outlook.

\section{Fundamentals of analytical CDFT}
\label{sec:fundamentals}

Before proceeding, we establish the notation used throughout: the indices $i, j, k, l, \dots, m, n, \dots$ denote occupied orbitals $a, b, c, d, \dots$ virtual orbitals, and $p, q, r, s, \dots$ general orbitals, while the index $f$ denotes a frontier orbital. Spin indices are represented by $\sigma, \tau, \lambda, \theta, \dots$, and the spin orbital energies are denoted by $\varepsilon_{p_\sigma}$. The foundation of conceptual density functional theory lies in the power series expansion of the energy functional with respect to its two intrinsic variables, namely the number of electrons $N$ and the external potential $v(\mathbf{r})$. In the canonical ensemble, this expansion reads

\begin{align}
E[N+\Delta N,v+\Delta v]&=E[N,v] + \left(\frac{\partial E}{\partial N}\right)_v\Delta N+\int\left(\frac{\delta E}{\delta v(\mathbf{r})}\right)_N \Delta v(\mathbf{r})d\mathbf{r} \notag \\
&\quad +\frac{1}{2}\left(\frac{\partial^2 E}{\partial N^2}\right)_v\Delta N^2+\Delta N\int\frac{\partial^2E}{\partial N\delta v(\mathbf{r})}\Delta v(\mathbf{r})d\mathbf{r} \notag \\
&\quad +\frac{1}{2}\int\int\left(\frac{\delta^2 E}{\delta v(\mathbf{r})\delta v(\mathbf{r}')}\right)_N\Delta v(\mathbf{r})\Delta v(\mathbf{r}')d\mathbf{r}d\mathbf{r}' + \cdots \notag \\
\Delta E[N,v]&= \mu\Delta N+\int\rho(\mathbf{r})\Delta v(\mathbf{r})d\mathbf{r}+\frac{1}{2}\eta\Delta N^2 \notag \\
&\quad +\Delta N\int f(\mathbf{r})\Delta v(\mathbf{r})d\mathbf{r}+\frac{1}{2}\int\int\chi(\mathbf{r},\mathbf{r}')\Delta v(\mathbf{r})\Delta v(\mathbf{r}')d\mathbf{r}d\mathbf{r}' + \cdots .
\end{align}

\noindent The energy derivatives or the response functions introduced in the above Taylor expansion are the CDFT reactivity indices. For example, the first-order descriptors are the chemical potential $\mu$\autocite{Parr1978} and the electron density $\rho(\mathbf{r})$;\autocite{Parr1989} the second-order descriptors are the chemical hardness $\eta$,\autocite{Parr1983,Pearson1963} the Fukui function $f(\mathbf{r})$,\autocite{Parr1984} and the linear response function $\chi(\mathbf{r},\mathbf{r}')$.\autocite{Parr1989} Higher-order response functions e.g., the third order dual descriptor\autocite{Morell2005} $f^{(2)}(\mathbf{r})$ and the Fukui kernel $g(\mathbf{r},\mathbf{r}')$,\autocite{Wang2026} (it is sometimes referred as $f(\mathbf{r,\mathbf{r}'})$ in literature, to name the CDFT indices systematically and avoid confusion with the Fukui matrix $F_{pq}$ we prefer denote it as $g(\mathbf{r},\mathbf{r}')$) capture chemical reactivity in greater detail. Two generalisations have recently been proposed to extend the applicability of the theory and to extract deeper physicochemical insight from it. First, for the ground state of electronic systems, the external potential $v(\mathbf{r})$ is replaced by the one-electron integrals $h_{pq}$, giving rise to conceptual density matrix functional theory (CDMFT).\autocite{Acke2023,Wang2025} The second one replaces the total number of electrons $N$ by the orbital occupation numbers $n_i$, leading to ground state orbital-based CDFT.\autocite{Liu1997} 
Extending this approach to excitated states through the generalised adiabatic connection Kohn--Sham (GAC-KS) theorem,\autocite{Gorling1999,Trushin2026,Wang2026} an arbitrary KS state index $J$ is introduced to the orbital CDFT in order to contain sufficient information about the excited state; the energy functional is then written as $E_J[\{n_{p_\sigma}\},v(\mathbf{r})]$, unifying the ground and the excited states description. An alternative approach to establish the foundation of the $\Delta$SCF approach, introduced by Yang and Ayers, is via the potential functional theory(PFT) as well as the $\Phi$ and $\gamma_S$ functional theories,\autocite{Yang2024} which also place the orbital CDFT for excited states on a firm ground. Putting everything together, the final generalised spin-resolved perturbation expansion of the energy function is then written as

\begin{align}
\Delta E_J[n_m,h_{pq}]&= \sum_m\left(\frac{\partial E_J}{\partial n_m}\right)_{\{h_{pq}\}}\Delta n_m+\sum_{pq}\left(\frac{\partial E_J}{\partial h_{pq}}\right)_{\{n_m\}} \Delta h_{pq} \notag \\
&\quad +\frac{1}{2}\sum_{mn}\left(\frac{\partial^2 E_J}{\partial n_m \partial n_n}\right)_{\{h_{pq}\}}\Delta n_m \Delta n_n+\sum_m\sum_{pq}\Delta n_m\frac{\partial^2E_J}{\partial n_m\partial h_{pq}}\Delta h_{pq} \notag \\
&\quad +\frac{1}{2}\sum_{pq}\sum_{rs}\left(\frac{\partial^2 E_J}{\partial h_{pq}\partial h_{rs}}\right)_{\{n_m\}}\Delta h_{pq}\Delta h_{rs} + \cdots  \notag \\
&= \sum_m\mu^J_m\Delta n_m + \sum_{pq}P_{pq}^J \Delta h_{pq} + \frac{1}{2}\sum_{mn}\eta_{mn}^J\Delta n_m \Delta n_n \notag \\
&\quad + \sum_m\sum_{pq}\Delta n_m F_{m,pq}^J\Delta h_{pq} + \frac{1}{2}\sum_{pq}\sum_{rs}X_{pqrs}^J\Delta h_{pq}\Delta h_{rs}+\cdots , \label{eq:expansion}
\end{align}

\noindent where the superscript $J$ for the newly introduced CDMFT descriptors will be dropped from now on for simplicity. By solving the spin-polarised coupled-perturbed self-consistent-field (CP-SCF) equations, analytical expressions for the newly introduced CDFT descriptors are obtained. Herein, we propose an original orbital CDMFT, in the most general form, where the first order contains the the (Kohn--Sham) orbital chemical potential $\mu_m$, which equals the KS orbital energy

\begin{align}
\mu_{m_\sigma} = \varepsilon_{m_\sigma} = \langle\phi_{m_\sigma}|\hat{h}|\phi_{m_\sigma}\rangle ,
\end{align}

\noindent and the density matrix $P_{pq}^{\sigma}$

\begin{align}
P_{pq}^\sigma = \sum_i^{\mathrm{occ}} n_{i_\sigma} \delta_{pi}\delta_{qi} .
\end{align}

\noindent The second order consists of the orbital hardness matrix $\eta_{m_\sigma n_\tau}$

\begin{align}
\eta_{m_\sigma n_\tau} &= K_{m_\sigma m_\sigma n_\tau n_\tau}^\eta - \sum_{\lambda\theta}\sum_{i_\lambda a_\lambda, j_\theta b_\theta} M^{-1}_{i_\lambda a_\lambda j_\theta b_\theta} K^{f}_{n_\tau n_\tau j_\theta b_\theta} K^{\eta}_{m_\sigma m_\sigma i_\lambda a_\lambda} \notag \\
&\quad - c_{\mathrm{HF}}\,\delta_{\sigma\tau}\sum_{p_\tau q_\tau} F_{n_\tau, p_\tau q_\tau}\,(\phi_{m_\sigma}\phi_{p_\tau}|\phi_{m_\sigma}\phi_{q_\tau}) ,
\label{eq:eta-orb}
\end{align}

\noindent which is also interpreted as the generalised screened interaction,\autocite{Mei2021} where the last term is the exact-exchange contribution for hybrid functionals ($c_{\mathrm{HF}}\neq 0$); the orbital Fukui matrix $F_{m_\sigma, p_\tau q_\tau}$ entering this term is

\begin{align}
F_{m_\sigma, p_\tau q_\tau} &= \frac{\partial P_{p_\tau q_\tau}}{\partial n_{m_\sigma}} \notag \\
&= \delta_{\sigma\tau}\delta_{m_\sigma q_\sigma}\delta_{m_\sigma p_\sigma} - \frac{1}{2}\sum_{\lambda}\sum_{i_\tau a_\tau, j_\lambda b_\lambda} M^{-1}_{i_\tau a_\tau j_\lambda b_\lambda} K^{f}_{m_\sigma m_\sigma j_\lambda b_\lambda} (\delta_{p_\sigma a_\sigma}\delta_{q_\sigma i_\sigma}+\delta_{q_\sigma a_\sigma}\delta_{p_\sigma i_\sigma}) ,
\label{eq:fukui_mat}
\end{align}

\noindent and the linear response matrix $X_{p_\sigma q_\sigma r_\tau s_\tau}$

\begin{align}
X_{p_\sigma q_\sigma, r_\tau s_\tau} &= \frac{\partial P_{p_\sigma q_\sigma}}{\partial h_{r_\tau s_\tau}} \notag \\
&= -\frac{1}{2} \sum_{i_\sigma a_\sigma, j_\tau b_\tau} M^{-1}_{i_\sigma a_\sigma j_\tau b_\tau} (\delta_{j_\tau r_\tau}\delta_{b_\tau s_\tau}+\delta_{j_\tau s_\tau}\delta_{b_\tau r_\tau})(\delta_{a_\sigma q_\sigma}\delta_{i_\sigma p_\sigma}+\delta_{a_\sigma p_\sigma}\delta_{i_\sigma q_\sigma}) .
\end{align}

\noindent The local version of the orbital Fukui function $f_{p_\sigma}(\mathbf{r})$ and linear response function $\chi(\mathbf{r},\mathbf{r'})$ are

\begin{align}
f_{p_\sigma}(\mathbf{r})&=|\phi_{p_\sigma}(\mathbf{r})|^2-\sum_{\lambda\tau}\sum_{i_\lambda a_\lambda j_\tau b_\tau}M_{i_\lambda a_\lambda j_\tau b_\tau}^{-1} K_{p_\sigma p_\sigma j_\tau b_\tau}^{f} \phi_{i_\lambda}(\mathbf{r})\phi_{a_\lambda}(\mathbf{r}) , \label{eq:fukui-r} \\
\chi(\mathbf{r},\mathbf{r}')&=-2\sum_{\sigma\tau}\sum_{i_\sigma a_\sigma j_\tau b_\tau}M_{i_\sigma a_\sigma j_\tau b_\tau}^{-1}\phi_{i_\sigma}(\mathbf{r})\phi_{a_\sigma}(\mathbf{r})\phi_{j_\tau}(\mathbf{r}')\phi_{b_\tau}(\mathbf{r}') , \label{eq:chi-rr}
\end{align}

\noindent where the matrices are defined in the preceding equations; these expressions hold when a single hybrid functional, e.g. B3LYP ($c_{\mathrm{HF}}=0.2$), is employed.

\begin{align}
M_{i_\sigma a_\sigma j_\tau b_\tau}&=\delta_{\sigma\tau}\delta_{ij}\delta_{ab}(\varepsilon_{a_\sigma}-\varepsilon_{i_\sigma})+2(\phi_{i_\sigma}\phi_{a_\sigma}|\phi_{j_\tau}\phi_{b_\tau}) \notag \\
&\quad -c_{\mathrm{HF}}\delta_{\sigma\tau}[(\phi_{i_\sigma}\phi_{b_\sigma}|\phi_{j_\sigma}\phi_{a_\sigma})+(\phi_{i_\sigma}\phi_{j_\sigma}|\phi_{a_\sigma}\phi_{b_\sigma})]+2(\phi_{i_\sigma}\phi_{a_\sigma}|f_{\mathrm{xc}}^{\sigma\tau}|\phi_{j_\tau}\phi_{b_\tau}) , \label{eq:Mdef} \\
K_{p_\sigma p_\sigma j_\tau b_\tau}^{f}&=2(\phi_{p_\sigma}\phi_{p_\sigma}|\phi_{j_\tau}\phi_{b_\tau})-2c_{\mathrm{HF}}\delta_{\sigma\tau}(\phi_{p_\sigma}\phi_{j_\sigma}|\phi_{p_\sigma}\phi_{b_\sigma})+2(\phi_{p_\sigma}\phi_{p_\sigma}|f_{\mathrm{xc}}^{\sigma\tau}|\phi_{j_\tau}\phi_{b_\tau}) , \\
K_{p_\sigma p_\sigma q_\tau q_\tau}^\eta &= (\phi_{p_\sigma}\phi_{p_\sigma}|\phi_{q_\tau}\phi_{q_\tau})+(\phi_{p_\sigma}\phi_{p_\sigma}|f_{\mathrm{xc}}^{\sigma\tau}|\phi_{q_\tau}\phi_{q_\tau}) , \\
K_{p_\sigma p_\sigma i_\lambda a_\lambda}^\eta &=(\phi_{p_\sigma}\phi_{p_\sigma}|\phi_{i_\lambda}\phi_{a_\lambda})+(\phi_{p_\sigma}\phi_{p_\sigma}|f_{\mathrm{xc}}^{\sigma\lambda}|\phi_{i_\lambda}\phi_{a_\lambda}) .
\end{align}

\begin{figure}[htbp]
\centering
\includegraphics[width=0.9\textwidth]{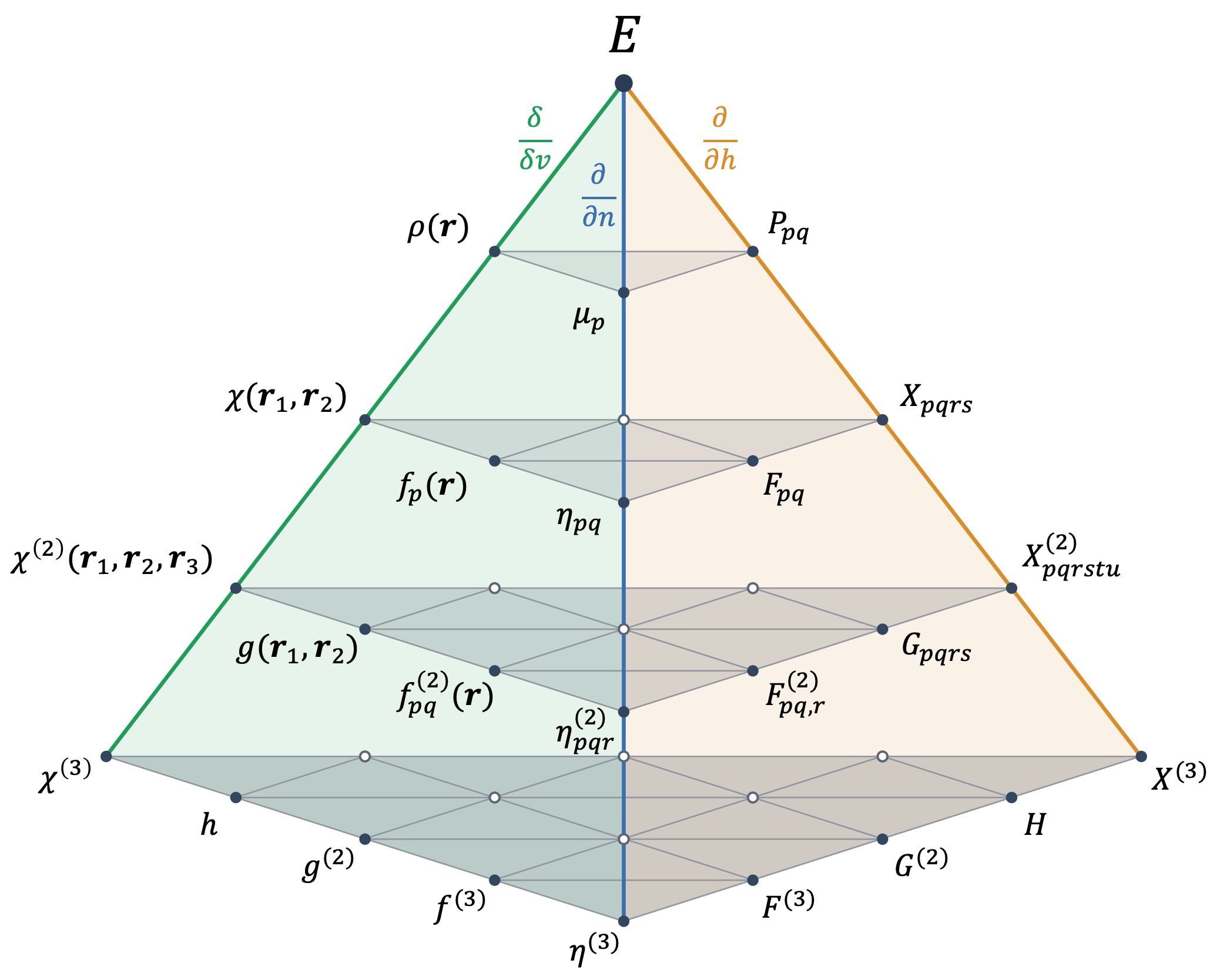}
\caption{Pyramid plot of the generalised orbital-based conceptual density (matrix) functional theory. The left green lateral face shows the orbital CDFT, while the orbital CDMFT is given on the right orange lateral face. Derivative level is indicated by shaded intersections. CDFT descriptors are marked by solide black dots, whereas irrelevant responses (mixed $v$ and $h$ derivates) are denoted by empty circles. The traditional CDFT is retrieved in the left green lateral face of the pyramid by replacing the $n$ by $N$ with concomitant disappearance of all indices.}
\label{fig:cdft_tetrahedron}
\end{figure}

As can be inferred from orbital conceptual density (matrix) functional theory, the $N$-derivatives of classical CDFT are a special case in which the derivative is taken at the frontier orbital occupation level, i.e. when $p=f$ so that $\mathrm{d}N=\mathrm{d}n_f$. A generalised CDFT pyramid plot is given in \cref{fig:cdft_tetrahedron}. The evaluation of the orbital CDFT for ground and excited states from \cref{eq:expansion} assumes the KS state $J$ is correctly mapped to the corresponding state of interest of the interacting electronic system e.g. $\Psi_I$ from the $\Delta$SCF approach. Thus the KS state index $J$ is inherited from the $\Delta$SCF wavefunction and the expansion in \cref{eq:expansion} evaluates the orbital CDFT response functions for that state with respect to the two independent variables $\{n_{m_\sigma}\}$ and $v(\mathbf{r})$.

\section{Development of PyCDFT}
\label{sec:development}

\subsection{Code structure and interoperability}

To evaluate the orbital CD(M)FT descriptors analytically and systematically, we developed \textsc{PyCDFT}, a pure Python library that computes them starting from a converged wavefunction produced by third-party software. The library is organized as a set of loosely coupled modules, each with a single responsibility: \texttt{fileio} collects the mean-field solution and writes all necessary information and settings into a master input file for the CDFT engine; \texttt{dataio} manages the creation, chunking, and compression of all intermediate and result datasets; \texttt{aim} performs the supported atoms-in-molecules (AIM) analyses and passes the resulting weights on for condensed CDFT evaluation; \texttt{helper} provides auxiliary functions that, e.g., classify the wavefunction, detect orbital degeneracies, and evaluate exchange--correlation kernels; and \texttt{acdft} implements the computational kernels that build, transform, and evaluate the response functions. A user-facing driver, \texttt{eval\_acdft}, chains these kernels into a complete workflow, but every stage can equally be invoked from an interactive Python session or a user script, e.g.\ to restart an error-terminated job, in keeping with the library-first philosophy of the modern Python simulation ecosystem. The largest share is borrowed from PySCF \autocite{Sun2018,Sun2020}, whose optimized C backends supply the two-electron integrals, DFT quadrature grids, and atomic-orbital values on grid points, so that PyCDFT inherits near-native integral performance without reimplementing this machinery. The exchange--correlation kernels $f_{\mathrm{xc}}$ are evaluated with Libxc \autocite{Lehtola2018}, tensor contractions and the coupled-perturbed solver rely on NumPy\autocite{Harris2020} and SciPy,\autocite{Virtanen2020} the MBIS partitioning is performed by \texttt{denspart}\autocite{Verstraelen2016}  with the \texttt{grid} library\autocite{Tehrani2024}, and parallelism and storage are handled by \texttt{mpi4py}\autocite{Dalcin2021} and \texttt{h5py} through the parallel HDF5 (MPI-IO) driver.

\begin{figure}[htbp]
\centering
\includegraphics[width=0.9\textwidth]{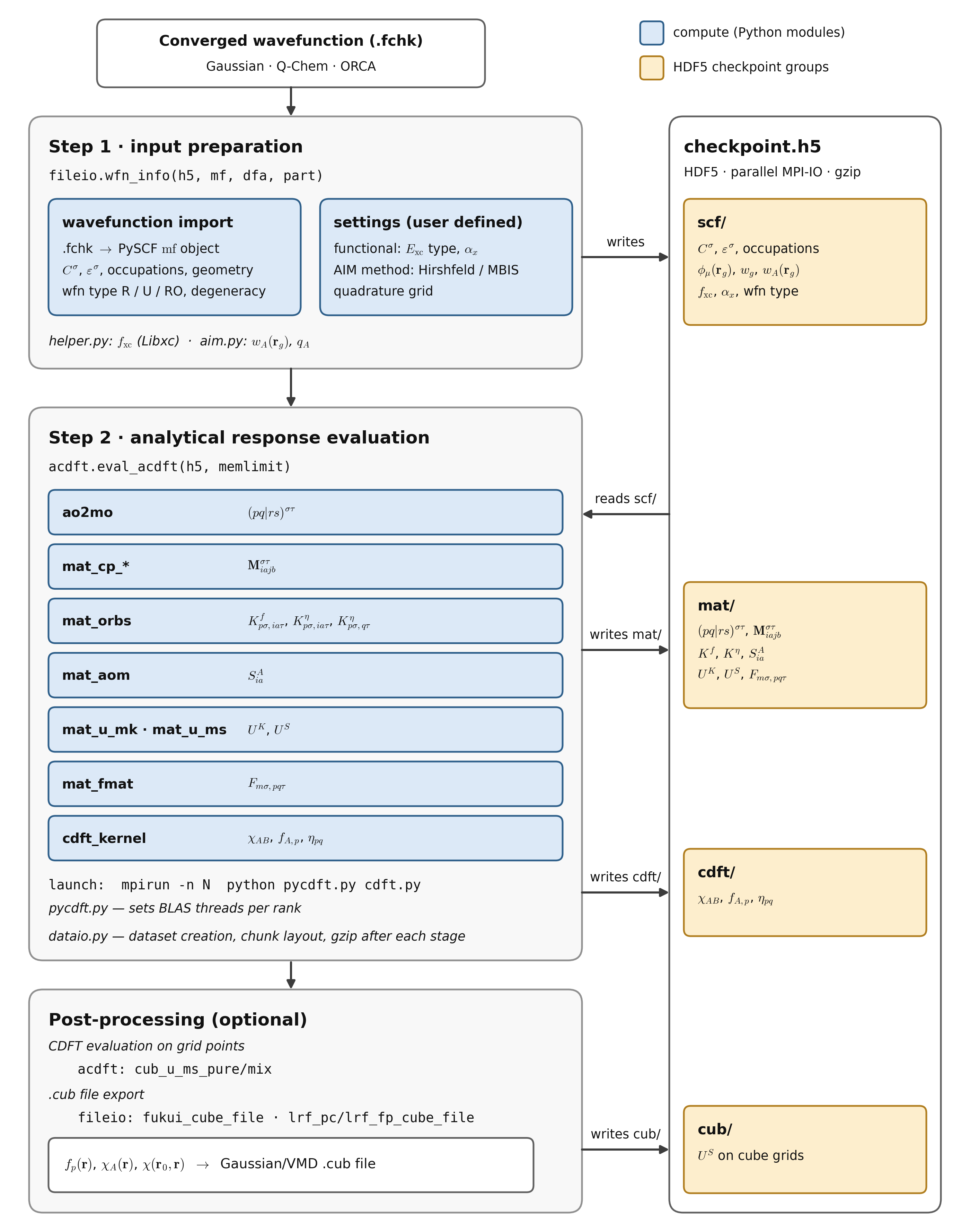}
\caption{Two-step workflow of PyCDFT: a converged wavefunction (.fchk) and user settings are written to a shared HDF5 checkpoint (Step 1), from which the response matrices and (condensed) CDFT descriptors are evaluated analytically (Step 2), with optional export of real-space descriptors as cube files.}
\label{fig:workflow}
\end{figure}

The workflow is summarized in Fig.~\ref{fig:workflow}. The reference Kohn--Sham DFT wavefunction may be converged with any electronic-structure package and is passed to PyCDFT as a Gaussian formatted checkpoint (fchk) file, written directly by Gaussian or Q-Chem, or generated from standard wavefunction formats (e.g., Molden) with tools such as MOKIT\autocite{mokit} or Multiwfn \autocite{Lu2012,Lu2024}. In the first step, the fchk data are loaded into a PySCF mean-field object and stored, together with the user-defined functional ($E_{\mathrm{xc}}$ type and $\alpha_x$), AIM scheme, and quadrature grid, in the \texttt{scf/} group of a single HDF5 checkpoint file that serves as the sole input for the second, computationally intensive step.

All expensive stages of the second step---integral generation, AO to MO transformation, coupled-perturbed matrix assembly, and the CP-SCF solutions---are parallelized by round-robin distribution of the outermost loop index over MPI ranks, which requires no communication beyond synchronization barriers because the work items are mutually independent. All ranks access the shared checkpoint concurrently through MPI-IO, writing datasets in slabs aligned with their chunk layout, and each finished dataset is repacked with gzip compression to reduce the disk footprint of the four-index quantities. Because each MPI rank additionally spawns its own set of BLAS threads, the parallel calculation is started through a lightweight launcher, \texttt{pycdft.py}, which fixes the number of threads per rank to the per-node core allocation before the numerical libraries are loaded; this prevents the thread oversubscription that would otherwise slow the compute stages several-fold on multi-core nodes. A calculation is therefore launched as \texttt{mpirun -np N python pycdft.py userjob.py}. The intermediate matrices reside in the \texttt{mat/} group---the transformed integrals $(pq|rs)^{\sigma\tau}$, the coupled-perturbed matrix $\mathbf{M}$, the coupling matrices $\mathbf{K}^{f}$ and $\mathbf{K}^{\eta}$, the atomic overlaps $\mathbf{S}$, the contracted response matrices $\mathbf{U}^{K}$ and $\mathbf{U}^{S}$, and the Fukui matrices $F_{m\sigma,pq\tau}$---while the condensed CDFT descriptors $\chi_{AB}$, $f_{A,p}$, and $\eta_{pq}$ are collected in \texttt{cdft/} and grid-resolved quantities for visualization in \texttt{cub/} if additional grid point evaluation is invoked (Sec.~\ref{sec:auxiliary}). The intermediates are retained on purpose, since CD(M)FT analyses reuse the $\mathbf{M}$ and $\mathbf{K}$ matrices and deleting them would force a costly recomputation; when they are no longer needed, the \texttt{mat/} group, by far the largest part of the file, can be safely removed with the provided file-reduction utility, leaving a compact record of the reference data and CDFT results.

\subsection{Implementation details}
\label{sec:implementation}

Although the working equations of Sec.~\ref{sec:fundamentals} are written in terms of the inverse coupled-perturbed matrix, $\mathbf{M}^{-1}$ is never constructed in practice. The dimension of $\mathbf{M}$ is $n_{\mathrm{occ}}n_{\mathrm{vir}}$ per spin channel, so an explicit inversion would scale as $\mathcal{O}(n_{\mathrm{occ}}^3 n_{\mathrm{vir}}^3)$ and return a dense matrix that is never needed as such: in every descriptor, $\mathbf{M}^{-1}$ appears contracted with a small set of vectors. PyCDFT therefore recasts each evaluation as a family of linear systems whose right-hand side is dictated by the target property,

\begin{align}
\sum_{\tau}\sum_{j_\tau b_\tau} M_{i_\sigma a_\sigma j_\tau b_\tau}\, U^{K}_{m_\lambda, j_\tau b_\tau} = K^{f}_{m_\lambda m_\lambda, i_\sigma a_\sigma} , \qquad
\sum_{\tau}\sum_{j_\tau b_\tau} M_{i_\sigma a_\sigma j_\tau b_\tau}\, U^{S}_{A, j_\tau b_\tau} = S^{A}_{i_\sigma a_\sigma} ,
\label{eq:mu-systems}
\end{align}

\noindent where $S^{A}_{i_\sigma a_\sigma}=\sum_g w_g\, w_A(\mathbf{r}_g)\,\phi_{i_\sigma}(\mathbf{r}_g)\phi_{a_\sigma}(\mathbf{r}_g)$ is the atomic overlap matrix of atom $A$ evaluated by quadrature with the AIM weights. The $\mathbf{MU}=\mathbf{K}$ family---one system per molecular orbital $m$---generates all orbital-resolved descriptors, whereas the $\mathbf{MU}=\mathbf{S}$ family---one system per atom $A$---generates the atom-condensed linear response function. Once the $\mathbf{U}$ matrices are known, all second-order descriptors reduce to inexpensive tensor contractions,

\begin{align}
\eta_{m_\sigma n_\tau} &= K^{\eta}_{m_\sigma m_\sigma n_\tau n_\tau} - \sum_{\lambda}\sum_{i_\lambda a_\lambda} K^{\eta}_{m_\sigma m_\sigma, i_\lambda a_\lambda}\, U^{K}_{n_\tau, i_\lambda a_\lambda} \notag \\
&\quad - c_{\mathrm{HF}}\,\delta_{\sigma\tau}\sum_{p_\tau q_\tau} F_{n_\tau, p_\tau q_\tau}\,(\phi_{m_\sigma}\phi_{p_\tau}|\phi_{m_\sigma}\phi_{q_\tau}) \\
f_{A,m_\sigma} &= S^{A}_{m_\sigma m_\sigma} - \sum_{\lambda}\sum_{i_\lambda a_\lambda} S^{A}_{i_\lambda a_\lambda}\, U^{K}_{m_\sigma, i_\lambda a_\lambda} , \\
\chi_{AB} &= -2 \sum_{\sigma}\sum_{i_\sigma a_\sigma} S^{B}_{i_\sigma a_\sigma}\, U^{S}_{A, i_\sigma a_\sigma} ,
\end{align}

\noindent where $S^{A}_{m_\sigma m_\sigma}$ is defined analogously to $S^{A}_{i_\sigma a_\sigma}$ with the orbital density $\phi_{m_\sigma}^2(\mathbf{r}_g)$. The last term of the hardness expression is the exact-exchange contribution, present only for hybrid functionals ($c_{\mathrm{HF}}\neq 0$) and only between same-spin orbitals; it is assembled from the orbital Fukui matrix $F_{n_\sigma,p_\sigma q_\sigma}$, whose off-diagonal blocks are built from $-\tfrac{1}{2}U^{K}_{n_\sigma}$, \cref{eq:fukui_mat}, so no additional linear systems are required. Note that only one of the two spin summations of Sec.~\ref{sec:fundamentals} is absorbed into the $\mathbf{U}$ matrices through \cref{eq:mu-systems}; the remaining sum over the spin of the occupied--virtual pair index must be carried out explicitly in the contractions.

The systems in \cref{eq:mu-systems} are solved by a direct or an iterative method, selected automatically from the system size. When $\mathbf{M}$ fits within the memory limit it is assembled once and factorized, so that every right-hand side follows by inexpensive back-substitution; larger systems fall back to a matrix-free, iterative solver. For a ground-state (Aufbau) reference $\mathbf{M}$ is symmetric positive definite, the regime in which the preconditioned conjugate gradient (CG) is the Krylov method of choice. Broken-symmetry and $\Delta$SCF references, however, generally yield an indefinite $\mathbf{M}$ for which CG has no convergence guarantee; PyCDFT detects this and switches automatically to the minimal residual method (MINRES), the corresponding method for symmetric indefinite systems, with an explicit residual check so that a solve which has not genuinely converged is never accepted. In the iterative path the matrix itself is never held in memory; it is exposed to SciPy as a matrix-free \texttt{LinearOperator} whose matrix--vector product streams slabs of $\mathbf{M}$ from the checkpoint file, so the memory footprint remains bounded by the user-defined limit irrespective of system size. Convergence is accelerated by a Jacobi preconditioner assembled once from the diagonal of the coupled-perturbed matrix,

\begin{align}
d_{i_\sigma a_\sigma} = M_{i_\sigma a_\sigma i_\sigma a_\sigma}
&= \varepsilon_{a_\sigma}-\varepsilon_{i_\sigma}
+ 2(\phi_{i_\sigma}\phi_{a_\sigma}|\phi_{i_\sigma}\phi_{a_\sigma})
+ 2(\phi_{i_\sigma}\phi_{a_\sigma}|f_{\mathrm{xc}}^{\sigma\sigma}|\phi_{i_\sigma}\phi_{a_\sigma}) \notag \\
&\quad - c_{\mathrm{HF}}[(\phi_{i_\sigma}\phi_{a_\sigma}|\phi_{i_\sigma}\phi_{a_\sigma})+(\phi_{i_\sigma}\phi_{i_\sigma}|\phi_{a_\sigma}\phi_{a_\sigma})] ,
\label{eq:jacobi}
\end{align}

\noindent taken for the operator actually solved, i.e., with the correspondingly spin-folded prefactors for a closed-shell reference (see below). The diagonal is dominated by the occupied--virtual gap $\varepsilon_a-\varepsilon_i$; because the off-diagonal coupling is comparatively weak, the preconditioned spectrum is strongly clustered and a few tens of iterations typically suffice at the default thresholds ($\texttt{rtol}=10^{-7}$, $\texttt{atol}=10^{-10}$). The essential structure of the solver, as implemented in \texttt{acdft}, is shown in \cref{lst:pcg}.

\begin{lstlisting}[style=pycode, caption={Essential structure of the matrix-free preconditioned conjugate-gradient solver for the $\mathbf{MU}=\mathbf{K}$ systems.}, label={lst:pcg}]
import numpy as np
from scipy.sparse.linalg import LinearOperator, cg

def matvec(x):
    """y = M @ x, with M streamed slab-wise from the checkpoint."""
    x = x.reshape(nocc, nvir)
    y = np.empty_like(x)
    for i in range(nocc):
        for a in range(nvir):
            y[i, a] = np.einsum('jb,jb->', M_h5[i, a], x)
    return y.ravel()

M_op = LinearOperator((nov, nov), matvec=matvec)
prec = LinearOperator((nov, nov), matvec=lambda r: r * dinv)  # dinv = 1/d

for m in range(nmo):          # independent RHS, round-robin over MPI ranks
    if m % size != rank:
        continue
    u, info = cg(M_op, K_f[m].ravel(), M=prec, rtol=1e-7, atol=1e-10)
    U_K[m] = u.reshape(nocc, nvir)
\end{lstlisting}

For a closed-shell reference the $\alpha$ and $\beta$ responses coincide, the systems are folded onto spatial orbital pairs of dimension $n_{\mathrm{occ}}n_{\mathrm{vir}}$, which doubles the Coulomb and exchange--correlation kernel prefactors of \cref{eq:Mdef} (the factors 2 become 4, with $f_{\mathrm{xc}}^{\sigma\tau}$ replaced by the total-density kernel) while the exact-exchange terms retain the single factor $c_{\mathrm{HF}}$. For an unrestricted reference the two spin channels remain coupled through the Coulomb and $f_{\mathrm{xc}}^{\alpha\beta}$ terms, and the same solver is applied to the block system

\begin{align}
\begin{pmatrix} \mathbf{M}^{\alpha\alpha} & \mathbf{M}^{\alpha\beta} \\[2pt] \mathbf{M}^{\beta\alpha} & \mathbf{M}^{\beta\beta} \end{pmatrix}
\begin{pmatrix} \mathbf{U}^{\alpha} \\[2pt] \mathbf{U}^{\beta} \end{pmatrix} =
\begin{pmatrix} \mathbf{R}^{\alpha} \\[2pt] \mathbf{R}^{\beta} \end{pmatrix}
\label{eq:block-system}
\end{align}

\noindent of dimension $n^{\alpha}_{\mathrm{occ}}n^{\alpha}_{\mathrm{vir}}+n^{\beta}_{\mathrm{occ}}n^{\beta}_{\mathrm{vir}}$, so that each orbital $m$ requires two solves, with stacked right-hand sides $(\mathbf{K}^{f,\alpha\alpha}_m,\mathbf{K}^{f,\alpha\beta}_m)$ and $(\mathbf{K}^{f,\beta\alpha}_m,\mathbf{K}^{f,\beta\beta}_m)$, and each atom one solve with $(\mathbf{S}^{A,\alpha},\mathbf{S}^{A,\beta})$. The Jacobi preconditioner generalizes trivially: the opposite-spin blocks contribute nothing to the diagonal of the block system, so $\mathbf{d}$ is the concatenation of the $\alpha\alpha$ and $\beta\beta$ diagonals of \cref{eq:jacobi}. To preserve the invariance of the descriptors under rotations within degenerate shells, the right-hand sides and the resulting descriptor rows are additionally averaged over degenerate orbital sets detected from the orbital energies. Since all right-hand sides share the same operator and are mutually independent, the solves are embarrassingly parallel and distributed round-robin over the MPI ranks.

The same machinery yields the real-space descriptors. Evaluating $f_{p_\sigma}(\mathbf{r})$ or $\chi(\mathbf{r},\mathbf{r}')$ of \cref{eq:fukui-r,eq:chi-rr} on a visualization grid requires the action of $\mathbf{M}^{-1}$ on the orbital-pair products $s_{ia}(\mathbf{r}_g)=\phi_{i}(\mathbf{r}_g)\phi_{a}(\mathbf{r}_g)$, so one $\mathbf{MU}=\mathbf{S}$ solve is performed per point $\mathbf{r}_g$ of the user-defined cube grid and the resulting $\mathbf{U}$ matrices are cached in the \texttt{cub/} group of the checkpoint file. Any kernel value $\chi(\mathbf{r}_g,\mathbf{r})$ is then synthesized as a simple contraction of the cached $\mathbf{U}$ matrices with orbital products, so that repeated visualizations at different iso-values or probe points incur no further solver cost (Sec.~\ref{sec:auxiliary}).

\subsection{Auxiliary functionality for analysis and visualization}
\label{sec:auxiliary}

Beyond the condensed descriptors, PyCDFT can evaluate the CDFT response functions on real-space grids for pictorial analysis. This cube analysis is carried out after the evaluation of the CDFT indices, so that all necessary matrices are already pre-computed and stored in the checkpoint file; one usually performs it at a later stage, once the condensed results have singled out the orbitals, atoms, or probe points worth visualizing. A built-in generator constructs a rectilinear cube grid from the molecular geometry, with a user-defined number of points along each axis and a padding distance around the atomic positions, and a companion writer exports any scalar field on this grid in the standard Gaussian cube format, so that iso-surfaces can be rendered directly in common viewers such as GaussView, VMD, or Avogadro. Three high-level functions cover the typical use cases. \texttt{fukui\_cube\_file} writes the orbital-resolved Fukui functions $f_{p_\sigma}(\mathbf{r})$ of \cref{eq:fukui-r} for a user-selected window of orbitals around the frontier. \texttt{lrf\_pc\_cube\_file} evaluates the partially condensed linear response function,

\begin{align}
\chi_A(\mathbf{r}) = -2 \sum_{\sigma}\sum_{i_\sigma a_\sigma} U^{S}_{A, i_\sigma a_\sigma}\, \phi_{i_\sigma}(\mathbf{r})\phi_{a_\sigma}(\mathbf{r}) ,
\label{eq:lrf-pc}
\end{align}

\noindent i.e., the response of the density at $\mathbf{r}$ to a perturbation condensed on atom $A$, obtained by contracting the stored $\mathbf{U}^{S}$ matrices of \cref{eq:mu-systems} with the orbital products on the cube grid. \texttt{lrf\_fp\_cube\_file} evaluates the fixed-point linear response function for an arbitrary probe point $\mathbf{r}_0$,

\begin{align}
\chi(\mathbf{r}_0,\mathbf{r}) = -2 \sum_{\sigma}\sum_{i_\sigma a_\sigma} U^{S}_{\mathbf{r}_0, i_\sigma a_\sigma}\, \phi_{i_\sigma}(\mathbf{r})\phi_{a_\sigma}(\mathbf{r}) ,
\label{eq:lrf-fp}
\end{align}

\noindent where $\mathbf{U}^{S}_{\mathbf{r}_0}$ is the solution of \cref{eq:mu-systems} with the point right-hand side $s_{i_\sigma a_\sigma}(\mathbf{r}_0)=\phi_{i_\sigma}(\mathbf{r}_0)\phi_{a_\sigma}(\mathbf{r}_0)$, taken from the $\mathbf{U}$ matrices stored in the \texttt{cub/} group (Sec.~\ref{sec:implementation}).

All of these functions operate solely on the checkpoint file as they are pure post-processing steps that require neither the original wavefunction file nor any recomputation of the response matrices, so grids can be refined, probe points moved, or orbital windows changed at negligible cost after the expensive evaluation has finished. For more specialized analyses, the underlying routines also return the raw two-point kernel $\chi(\mathbf{r},\mathbf{r}')$ on arbitrary user-supplied point sets, which can be processed further in Python, e.g., to integrate the kernel over custom regions or to construct domain-condensed variants beyond the built-in AIM schemes.

\section{Examples}
\label{sec:examples}

\subsection{Ground state of H$_2$O}
\label{sec:example-h2o}

As a first example, we evaluate the complete set of second-order CDFT descriptors for ground-state H$_2$O. The reference wavefunction is converged in Gaussian at the PBE level and exported as a formatted checkpoint file (\texttt{H2O\_gs.fch}); \cref{lst:h2o-genh5,lst:h2o-cdft} then constitute the entire PyCDFT input. The molecule is reconstructed from the fchk file with MOKIT, and a restricted Kohn--Sham object is built merely as a container: a single SCF cycle initializes the quadrature grid and the checkpoint file, after which the molecular-orbital coefficients and energies are overwritten by the converged values read from the fchk file. The density functional approximation is specified as a list \texttt{[exchange, correlation, family, $c_{\mathrm{HF}}$]} using Libxc functional names---here PBE exchange and correlation with $c_{\mathrm{HF}}=0$, consistent with the pure GGA reference calculation. The calculation is organized in two short scripts that mirror the two steps of Fig.~\ref{fig:workflow}. The first, \texttt{genh5.py}, prepares the HDF5 checkpoint input; since it creates the file with the serial HDF5 driver, it is executed as an ordinary serial Python script (\cref{lst:h2o-genh5}).

\begin{lstlisting}[style=pycode, caption={\texttt{genh5.py}: preparation of the HDF5 checkpoint input for ground-state H$_2$O, executed serially as \texttt{python genh5.py}.}, label={lst:h2o-genh5}]
from pyscf import dft
from mokit.lib.gaussian import load_mol_from_fch, mo_fch2py
from fileio import read_mo_energy, wfn_info

h5name  = 'H2O_gs.h5'
fchname = 'H2O_gs.fch'

# density functional approximation: [exchange, correlation, family, c_HF]
dfa = ['GGA_X_PBE', 'GGA_C_PBE', 'GGA', 0.0]       # PBE

# rebuild the molecule and a mean-field container from the fchk file
mol = load_mol_from_fch(fchname)
mol.verbose = 3
mf = dft.RKS(mol)
mf.max_cycle = 1                 # container only: orbitals replaced below
mf.chkfile = h5name
mf.grids.atom_grid = (75, 434)
mf.kernel()

# overwrite with the converged wavefunction from the fchk file
# MO occupation
# default Aufbau principle
# MO coefficient
mf.mo_coeff  = mo_fch2py(fchname)
# MO energy
mf.mo_energy = read_mo_energy(fchname)

# write the scf/ group of the checkpoint file
wfn_info(h5name, mf, dfa, part='MBIS')
\end{lstlisting}

The second script, \texttt{cdft.py}, performs the computationally intensive evaluation. It reads its entire input from the checkpoint file, fills the \texttt{mat/} and \texttt{cdft/} groups, and is launched under MPI through the \texttt{pycdft.py} launcher (Sec.~\ref{sec:development}), with the work distributed over the ranks as described in Sec.~\ref{sec:implementation} (\cref{lst:h2o-cdft}).

\begin{lstlisting}[style=pycode, caption={\texttt{cdft.py}: parallel analytical CDFT evaluation, executed as \texttt{mpirun -np 8 python pycdft.py cdft.py}.}, label={lst:h2o-cdft}]
from acdft import eval_acdft

eval_acdft('H2O_gs.h5', memlimit=1024)   # memory limit in MB per process
\end{lstlisting}

\noindent Upon normal termination, the condensed descriptors $\chi_{AB}$, $f_{A,p}$, and $\eta_{pq}$ reside in the \texttt{cdft/} group of \texttt{H2O\_gs.h5} and can be inspected with a few lines of \texttt{h5py}. One can write a simple script to extract the condensed orbital CDFT response functions as well as the traditional CDFT indices from the checkpoint as follows.

\begin{lstlisting}[style=pycode, caption={Extrating data from the h5 checkpoint file.}, label={lst:data-ext}]
import h5py

with h5py.File('H2O_gs.h5', 'r') as f:
    # orbital CDFT descriptors
    lrf   = f['cdft/lrf_a'][...]     # condensed linear response chi_AB
    fukui = f['cdft/ff_a'][...]      # condensed orbital Fukui f_{A,p}
    eta   = f['cdft/eta_aa'][...]    # orbital hardness eta_pq
    # orbital indices
    occ   = f['scf/occ_a'][...]      # occupied MO indices
    vir   = f['scf/vir_a'][...]      # virtual  MO indices

# traditional CDFT: the frontier limit p = f, i.e. dN = dn_f
homo, lumo = occ[-1], vir[0]    # HOMO, LUMO indices
fukui_minus = fukui[:, homo]    # condensed Fukui f_A^-
fukui_plus  = fukui[:, lumo]    # condensed Fukui f_A^+
eta_minus   = eta[homo, homo]   # hardness eta^-
eta_plus    = eta[lumo, lumo]   # hardness eta^+
\end{lstlisting}

\subsection{Singly excited state of NH$_3$ via $\Delta$SCF}
\label{sec:example-nh3}

The second example demonstrates that PyCDFT is not limited to ground states: any single-determinant reference can be analyzed, including $\Delta$SCF solutions for excited states. Here the lowest singly excited singlet state of NH$_3$ is converged in Q-Chem with the restricted open-shell Kohn--Sham (ROKS) approach at the PBE0 level, and the converged wavefunction is exported to the fchk format. A broken-symmetry unrestricted determinant is then constructed from the ROKS orbitals to represent the singlet excited-state density and its response. The preparation script (\cref{lst:nh3-genh5}) accordingly differs from that of Sec.~\ref{sec:example-h2o} in three respects. First, an unrestricted container (\texttt{UKS}) is used, and the occupation vector is edited by hand to the open-shell singly excited configuration---one $\beta$ electron is moved from the HOMO to the LUMO---so that the broken-symmetry reference is presented to PyCDFT as an unrestricted wavefunction and the spin-resolved machinery of Sec.~\ref{sec:implementation} is engaged automatically. Second, the hybrid functional is specified through its combined Libxc name (\texttt{HYB\_GGA\_XC\_PBEH}, with the correlation entry set to \texttt{None}), so that the semilocal part of the kernel $f_{\mathrm{xc}}$ is returned with the correct internal scaling, while $c_{\mathrm{HF}}=0.25$ supplies the exact-exchange fraction entering \cref{eq:Mdef}. Third, the condensed descriptors are requested with the Hirshfeld partitioning, for which the spherically averaged pro-atom densities are generated on the fly from neutral-atom calculations with the same functional and basis set.

\begin{lstlisting}[style=pycode, caption={\texttt{genh5.py}: preparation of the checkpoint input for the broken-symmetry singly excited state of NH$_3$, executed serially as \texttt{python genh5.py}.}, label={lst:nh3-genh5}]
from pyscf import dft
from mokit.lib.gaussian import load_mol_from_fch, mo_fch2py
from fileio import read_mo_energy, wfn_info

h5name  = 'NH3_bs.h5'                 # bs: broken-symmetry reference
fchname = 'NH3_bs.fch'

# hybrid functional via its combined Libxc name: PBE0
dfa = ['HYB_GGA_XC_PBEH', None, 'GGA', 0.25]

# rebuild the molecule and a mean-field container from the fchk file
mol = load_mol_from_fch(fchname)
mol.verbose = 3
mf = dft.UKS(mol)
mf.max_cycle = 1                   # container only: orbitals replaced below
mf.chkfile = h5name
mf.grids.atom_grid = (75, 434)
mf.kernel()

# singly excited configuration: promote one beta electron (HOMO -> LUMO)
mf.mo_occ[1][4] = 0
mf.mo_occ[1][5] = 1

# overwrite with the converged ROKS Delta-SCF wavefunction from the fchk file
mf.mo_coeff  = mo_fch2py(fchname)
mf.mo_energy = read_mo_energy(fchname)

# write the scf/ group of the checkpoint file
wfn_info(h5name, mf, dfa, part='Hirshfeld')
\end{lstlisting}

The evaluation script is identical to \cref{lst:h2o-cdft} and is not repeated. Because the reference is open-shell, the second step solves the coupled block system of \cref{eq:block-system}, and the \texttt{cdft/} group is populated with the spin-resolved descriptors, i.e., $\chi^{\sigma}_{AB}$, $f^{\sigma}_{A,p}$, and the four spin blocks $\eta^{\sigma\tau}_{pq}$ of the orbital hardness.


\subsection{Cube files for iso-surface visualization}
\label{sec:example-cube}

The final example illustrates the auxiliary functionality of Sec.~\ref{sec:auxiliary}, using the checkpoint file produced in Sec.~\ref{sec:example-h2o}. Since all required matrices are already stored, this is a pure post-processing step executed as a short serial script at any time after the evaluation. Each function generates the cube grid internally from the molecular geometry, with \texttt{grid\_point} voxels along each axis and a padding of \texttt{delta} bohr around the atomic positions, and writes standard \texttt{.cub} files that can be opened directly in GaussView or VMD. In \cref{lst:cube}, the partially condensed linear response function $\chi_A(\mathbf{r})$ of \cref{eq:lrf-pc} is exported for the first two atoms (atom indices are zero-based), the fixed-point linear response function $\chi(\mathbf{r}_0,\mathbf{r})$ of \cref{eq:lrf-fp} is evaluated for a perturbation placed at the origin, drawing on the $\mathbf{U}$ matrices stored in the \texttt{cub/} group, and the orbital Fukui functions $f_{p}(\mathbf{r})$ of \cref{eq:fukui-r} are written for a frontier window of \texttt{window}$+1$ orbitals on each side of the gap, here HOMO$-$1, HOMO, LUMO, and LUMO$+$1.

\begin{lstlisting}[style=pycode, caption={\texttt{cube.py}: post-processing export of real-space CDFT descriptors as Gaussian cube files, executed as \texttt{python cube.py}.}, label={lst:cube}]
import numpy as np
import fileio

filename = 'H2O_gs.h5'

# partially condensed LRF chi_A(r) for the first two atoms (O and H)
fileio.lrf_pc_cube_file(filename, np.array([0, 1]), grid_point=(50, 50, 50), delta=5)

# fixed-point LRF chi(r0, r) with the perturbation placed at the origin
fileio.lrf_fp_cube_file(filename, np.array([0.0, 0.0, 0.0]))

# orbital Fukui functions f_p(r) for HOMO-1, HOMO, LUMO, LUMO+1
fileio.fukui_cube_file(filename, window=1, grid_point=(50, 50, 50), delta=5)
\end{lstlisting}


\section{Conclusions}
\label{sec:conclusions}

We have presented \textsc{PyCDFT}, a Python-scriptable library for the analytical evaluation of conceptual density (matrix) functional theory descriptors. From a single converged wavefunction, imported from essentially any electronic-structure package, the library evaluates the ground and excited states' orbital-resolved and atom-condensed second-order descriptors---orbital hardness, Fukui functions, and the linear response function---without recourse to finite-difference protocols, for closed-shell as well as broken-symmetry open-shell references, and exports real-space descriptors for direct visualization. The worked examples show that a complete analysis requires only a few lines of user code. Beyond its immediate use in chemical reactivity studies, the analytical machinery opens several directions. The exact descriptors provide reference values against which the common approximate CDFT schemes can be benchmarked systematically, and the response quantities themselves may serve in density-based method development, e.g., in the construction and diagnosis of density functionals or as features for machine-learned reactivity models. On the code side, planned improvements include a hierarchy of cheaper approximations to the coupling matrix $\mathbf{M}$ for large systems, GGA and meta-GGA kernels,\autocite{Wang2023} additional AIM partitionings, and extensions toward higher-order and field-dependent descriptors of conceptual density matrix functional theory; given its modular, library-first design, the functionality could also be merged into widely used packages such as PySCF or Psi4, bringing analytical CDFT to their broad user communities. We hope that a standardized, open implementation will lower the entry barrier for analytical CDFT and make new reactivity concepts easy to test against a common reference.

\section*{Acknowledgement}

FDP acknowledges support of the Vrije Universiteit Brussel through a Strategic Research Program (SRP73) awarded to his research group. BW acknowledges the junior postdoctoral fellowship support from FWO research project grant 1225526N.

\section*{Data Availability}

\textsc{PyCDFT} is free and open-source software, licensed under the Apache License, Version 2.0. Source code, installation instructions, worked examples with reference data, and detailed documentation are available at \url{https://github.com/lucienwb/PyCDFT}.

\printbibliography

\newpage

\begin{figure}[htbp]
\centering
\includegraphics[width=0.9\textwidth]{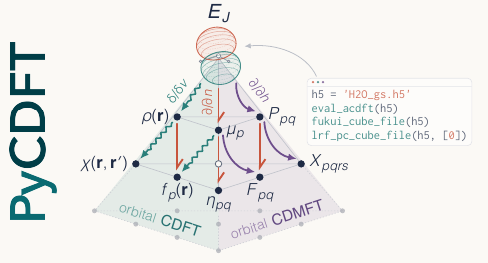}
\caption*{Graphical abstract}
\end{figure}

\end{document}